\pdfoutput=1
\documentclass[aps,prl,floatfix,twocolumn,reprint]{revtex4-1}
\usepackage{physics}
\usepackage{amsfonts}
\usepackage{mathrsfs}
\usepackage{amsmath}
\usepackage{color}
\usepackage{graphicx}
\usepackage{bm}
\usepackage{amssymb}
\usepackage{xspace}
\usepackage{epstopdf}
\usepackage{dcolumn}
\usepackage{longtable}
\usepackage[colorlinks=true, letterpaper=true, pdfstartview=FitV, linkcolor=blue, citecolor=blue, urlcolor=blue]{hyperref}
\begin{document}
\title{Topological Majorana Two-Channel Kondo Effect}
\author{Zhi-qiang Bao}
\author{Fan Zhang}\email{zhang@utdallas.edu}
\affiliation{Department of Physics, University of Texas at Dallas, Richardson, Texas 75080, USA}
\begin{abstract}
A one-dimensional time-reversal-invariant topological superconductor
hosts a Majorana Kramers pair at each end,
where time-reversal symmetry acts as a supersymmetry that flips local fermion parity.
We examine the transport anomaly of such a superconductor, floating and tunnel-coupled to normal leads at its two ends.
We demonstrate the realization of
a topologically-protected, channel-symmetric, two-channel Kondo effect without fine-tuning.
Whereas the nonlocal teleportation vanishes, 
a lead present at one end telecontrols the universal transport through the other end.
\end{abstract}
\maketitle

\indent\textcolor{blue}{\em Introduction.}---One of the current priorities in physics is to create a topological superconductor (TSC) 
that hosts Majorana bound states~\cite{Read,Kitaev}, which are their own antiparticles. 
Aside from their fundamental interest, Majoranas may offer a decoherence-free method 
for storing and manipulating quantum information~\cite{Nayak}. 
For spinful electrons, there are two classes of TSCs, D and DIII~\cite{Schnyder,Kitaev2,Chiu}.
In class D, time-reversal ($\mathbb{T}$) symmetry is broken;
one spin species is gapped out whereas the other effectively forms a 
$p$-wave superconductor (SC)~\cite{Kane1,Z,T,S,A,L,O,Alicea,Beenakker,Stanescu}.
It follows that an unpaired Majorana appears at a boundary.

\begin{figure}[!b]
\includegraphics[width=1.0\columnwidth]{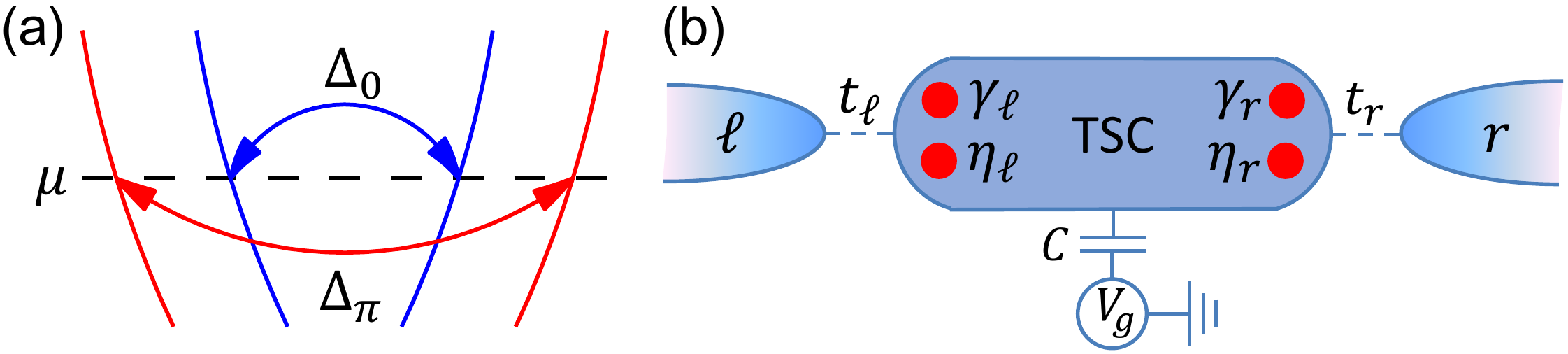}
\caption{(a) Illustration of realizing a 1D $\mathbb{T}$-invariant TSC:
if the order parameter $\Delta_{0,\pi}$ switches sign 
from the inner pair of bands to the outer pair,
there emerges a Majorana Kramers pair $\gamma$ and $\eta$ at each end of the TSC.
(b) Schematics of a Majorana transistor: the $\mathbb{T}$-invariant TSC 
grounded by a capacitor and tunnel-coupled to the left and right leads.}
\label{fig1}
\end{figure}

In class DIII, which will be the focus hereafter, $\mathbb{T}$ symmetry is respected and crucial.
SC order parameters can be viewed as mass terms that gap the Fermi surfaces.
Analogous to the band inversions in $\mathbb{Z}_2$ topological insulators (TI)~\cite{Kane,Qi},
a 1D TSC can be achieved once the order parameter switches sign at odd pairs of bands~\cite{Qi2,FZ1}.
Remarkably on each end, there emerges one Kramers pair of Majoranas.  
The simplest case with two pairs of bands is illustrated in Fig.~\ref{fig1}(a).
Crucially, in SCs quasiparticles must come in particle-hole pairs,
whereas $\mathbb{T}$ symmetry dictates them to come in Kramers pairs.
Therefore, the Majorana Kramers pair is topologically protected as a whole; 
the pair can neither split in energy nor move away from zero energy,
unless there is a bulk phase transition with gap closure.
The Majorana Kramers pairs also enjoy symmetry-protected non-Abelian braiding statistics~\cite{NA1,NA2}.
Different schemes have been proposed to realize 1D $\mathbb{T}$-invariant TSCs,
following a no-go theorem proposed by Zhang~\cite{talk}. 
Fig.~\ref{fig1}(a) can be effectively achieved 
in an $s$-wave Josephson $\pi$-junction mediated by the helical edge state of 2D TI~\cite{Kane2,Berg,Schrade,FZ3,Lee}.
This scheme has been under active experimental study~\cite{Knez,Hart,Shi,Pri,Boc}.
In a Rashba nanowire, the $s_{\pm}$-wave pairing in Fig.~\ref{fig1}(a) can be proximity induced by 
a nodeless iron-based SC~\cite{FZ1}, or other unconventional ones~\cite{Wong,Nakosai}.
This scheme constitutes an advantage of enjoying a higher $T_c$~\cite{Mazin,Bozovic}.
 Electron-electron interactions~\cite{Flen,Loss,Haim} have also been utilized in nanowire systems 
 to drive similar $s_{\pm}$-wave pairing.
 
As $\mathbb{T}^{2}$$=$$-1$ for spinful electrons, 
upon successive $\mathbb{T}$ the two Majoranas must interchange as follows
\begin{eqnarray}
\mathbb{T}:\quad \gamma_{\alpha}\rightarrow\eta_{\alpha}\rightarrow-\gamma_{\alpha}\,,
\label{TM}
\end{eqnarray}
where $\alpha$ labels the left ($\ell$) and right ($r$) ends of the TSC.
As two Majoranas form a Dirac fermion, one can define 
$d_{\alpha}$$=$$(\gamma_{\alpha}$$+$$i\eta_{\alpha})/2$ and $n_{\alpha}$$=$$d_{\alpha}^\dag d_{\alpha}$ 
as the annihilation and number operators of the corresponding complex fermion levels.
It follows from Eq.~(\ref{TM}) that
\begin{eqnarray}
\mathbb{T}:\quad d_{\alpha}\rightarrow id_{\alpha}^\dag\,,\quad 
d_{\alpha}^\dag\rightarrow -id_{\alpha}\,,\quad 
n_{\alpha}\rightarrow 1-n_{\alpha}\,.
\label{TF}
\end{eqnarray}  
This implies that for each $d$-level 
the only two accessible states with it being occupied or empty are Kramers partners.
$d_{\alpha}$ ($i d_{\alpha}^\dag$) may be viewed
as $d_{\alpha\uparrow}$ ($d_{\alpha\downarrow}$), reflecting a particle-hole redundancy.
Appealingly, $\mathbb{T}$ symmetry acts  as a local supersymmetry, 
not only flipping spin but also switching fermion parity~\cite{SUSY}.

If the TSC is mesoscopic and grounded by a capacitor, as sketched in Fig.~\ref{fig1}(b), 
it acquires a charging energy~\cite{Charging}
\begin{eqnarray}\label{Charging}
\mathcal{U}_Q=\frac{(Q-Q_0)^2}{2C}\quad\mbox{with}\quad Q=(n_\ell+n_r+2n_c)e\,.
\end{eqnarray}
Here $C$ is a capacitance, $Q_0$ is a gate charge, 
and both are controllable by the capacitor and the gate across it~\cite{Marcus}. 
$n_c$ is the number of Cooper pairs in the condensate. 
As a result, states at different fillings of the supersymmetric levels are no longer degenerate. 
We focus on a general case in which $e^2/C$ is several times smaller than the TSC gap $\Delta$ 
and $Q_0/e$ is rounded to $2N$$+$$1$.
Thus, the many-body state of TSC $(n_\ell, n_r, n_c)$ must be  
either $(1,0,N)$ or $(0,1,N)$, which are Kramers degenerate.
Such a floating TSC is analogous to a quantum dot (QD) with a singly occupied level, 
yet the Kramers degeneracy of TSC has a topological origin.
When the QD is weakly probed by the source and drain,
a one-channel Kondo (1CK) effect occurs in such a single-electron transistor~\cite{GR,NL,Kastner,KG}.
One might naturally wonder whether the floating TSC could mediate a similar 1CK effect.
Remarkably enough, such a Majorana transistor exhibits an unusual channel-symmetric two-channel Kondo (2CK) effect instead.          

The Kondo physics provides a paradigm for scaling and universality.
In the 1CK effect~\cite{GR,NL,Kastner,KG}, a twofold degeneracy such as a local spin-$\frac{1}{2}$ impurity is 
antiferromagnetically coupled to a reservoir of conduction electrons. 
The ground state, although complex, is a Fermi liquid in which the local spin is exactly screened 
and the low-lying excitations are quasiparticles.
In the more tantalizing 2CK effect~\cite{Nozieres,Zawadowski,Affleck}, 
two independent reservoirs compete to screen the spin and thus overscreen it, 
leaving an effective spin-$\frac{1}{2}$ impurity to be overscreened in the next stage. 
In such a ``domino'' effect, there is an emergent spin at each stage, 
and all electrons are strongly frustrated, leading to non-Fermi-liquid quantum criticality.
The delicate 2CK effect is unlikely to occur, 
as any channel asymmetry relieves the frustration and 
drives the system toward the 1CK effect in the more strongly coupled channel.
Realizing the 2CK effect without fine-tuning the channel symmetry has long been an unrealized dream.
Recently, three experiments have provided compelling evidences 
for spin, charge, and orbital 2CK effects in highly fine-tuned double QDs~\cite{ex1}, 
$\nu\!=\!1$ quantum Hall liquid~\cite{ex2}, and ferromagnetic $L1_0$-MnAl~\cite{ex3}, respectively. 
In this Letter, we demonstrate the realization of a topologically protected, channel-symmetric, 
2CK effect without fine-tuning in the Majorana transistor. 
Surprisingly, whereas the nonlocal electron teleportation vanishes, 
a lead coupled to one end of the floating TSC telecontrols 
the universal electron transport through the other end.

\indent\textcolor{blue}{\em Model and RG analysis.}---We start by considering a floating TSC coupled to 
two normal leads through tunneling, as sketched in Fig.~\ref{fig1}(b).
Analogous to the single-electron transistor made from a QD,
such a ``Majorana transistor" can be described by
\begin{eqnarray}\begin{aligned}
\mathcal{H}=&\sum_{\bm{k}\alpha\sigma}\varepsilon_{\bm{k}\alpha\sigma}
c_{\bm{k}\alpha\sigma}^{\dagger}c_{\bm{k}\alpha\sigma}+\mathcal{U}_Q  \\
+&\sum_{\bm{k}\alpha}t_{\alpha}(c_{\bm{k}\alpha\uparrow}d_{\alpha}^{\dagger}
-ic_{\bm{k}\alpha\downarrow}d_{\alpha}e^{i\phi}+\mbox{H.c.})\,,
\label{HT}
\end{aligned}\end{eqnarray}
where $c_{\bm{k}\alpha\sigma}$ is the annihilation operator of an electron of energy 
$\varepsilon_{\bm{k}\alpha\sigma}$ and spin $\sigma$ ($\uparrow,\downarrow$) in the lead $\alpha$ ($\ell,r$).
The tunneling matrix elements $t_\alpha$ are real, spin-momentum independent, and much smaller than $e^2/C$.
In Eq.~(\ref{HT}) the two tunneling terms are related by $\mathbb{T}$ symmetry.
The factor $e^{i\phi}$ creates one Cooper pair in the condensate
in a tunneling event mediated by Majoranas, 
given that $\phi$ is the SC phase and that $[Q,\phi]\!=\!2ei$.
The appearance of phase factors only in those events participated 
by lead electrons of $\sigma\!=\,\downarrow$ simply reflects a gauge choice.

\begin{figure}[!t]
\includegraphics[width=1.0\columnwidth]{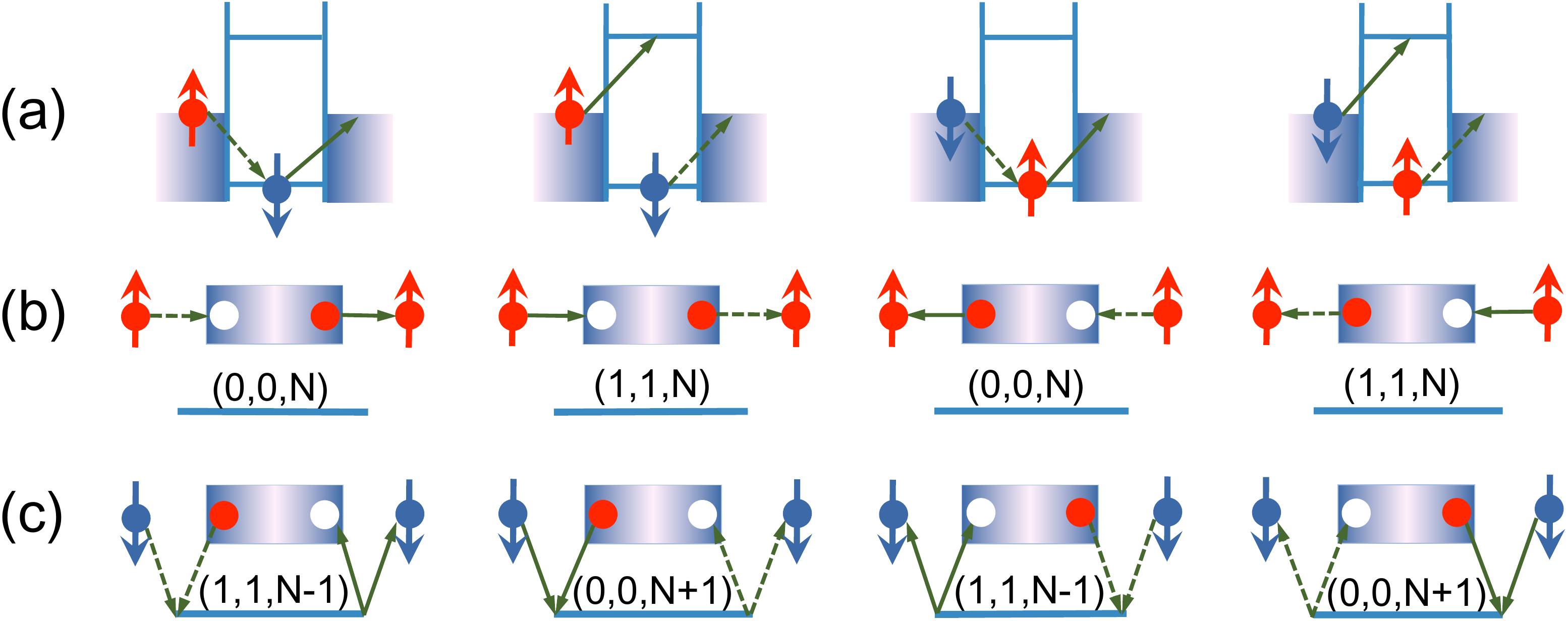}
\caption{Illustration of 2CK physics in Majorana transistor. 
(a) The co-tunneling events flipping spin in the QD.
A solid (dashed) arrow denotes the earlier (later) tunneling in a co-tunneling event.
(b)-(c) The co-tunneling events in the $\uparrow$ and $\downarrow$ channels of the Majorana transistor.
The events in (b) one-to-one correspond to those in (a).
The events in (b) and (c) are one-to-one switched under $\mathbb{T}$;
a white (red) dot denotes an empty (filled) $d$-level;
a horizontal blue line denotes the condensate;
$(n_\ell,n_r,n_c)$ denotes a virtual state.}
\label{fig2}
\end{figure}

Since we focus on the many-body states with $Q\!=\!2N+1$ 
that minimize the charging energy $\mathcal{U}_Q$,
the concerned odd-parity subspace must be spanned by the Kramers partners $(1,0,N)$ and $(0,1,N)$, 
as motivated in the {\em Introduction}.
First order tunneling events are strongly suppressed, 
as they lead to states with higher charging energies.
In contrast, the second order co-tunneling events are dominating,
as they yield states within the subspace of interest.
Hence, we perform the standard Schrieffer-Wolff transformation~\cite{SWT} to
project Eq.~(\ref{HT}) into the concerned subspace up to second order in $t_\alpha$.
This gives rise to a Kondo $s$-$d$ exchange model:
\begin{eqnarray}\begin{aligned}
\mathcal{H}_{\rm 2CK}=&\sum_{\bm{k}\alpha\sigma}\varepsilon_{\bm{k}\alpha\sigma}
c_{\bm{k}\alpha\sigma}^{\dagger}c_{\bm{k}\alpha\sigma} 
+\sum_{\bm{k}\bm{k}'\alpha\sigma}\frac{J_\delta}{2}
\sigma c_{\bm{k}\alpha\sigma}^{\dagger}c_{\bm{k}'\alpha\sigma}S_{z}\\
+&\sum_{\bm{k}\bm{k}'\sigma} \big[ J_{z}s_{z\sigma}S_{z} 
+\frac{J_{+}}{2}s_{-\sigma}S_{+}+\frac{J_{-}}{2}s_{+\sigma}S_{-} \big]\,,
\label{H2CK}
\end{aligned}\end{eqnarray}
where the spin $\bm{S}$ of the floating TSC and
the pseudospin $\bm{s}_\sigma$ of the lead electrons 
are defined as
\begin{equation}\begin{split}
2S_{z}=d_{\ell}^{\dagger}d_{\ell}-d_{r}^{\dagger}d_{r},&\quad
S_{+}=S_{-}^{\dagger}=d_{\ell}^{\dagger}d_{r},\\
2s_{z\uparrow}\!=\!c_{\bm{k}\ell\uparrow}^{\dagger}c_{\bm{k}'\ell\uparrow}
\!-\!c_{\bm{k}r\uparrow}^{\dagger}c_{\bm{k}'r\uparrow},&\quad
s_{+\uparrow}\!=\!s_{-\uparrow}^{\dagger}\!=\!c_{\bm{k}\ell\uparrow}^{\dagger}c_{\bm{k}'r\uparrow},\\
2s_{z\downarrow}\!=\!c_{\bm{k}r\downarrow}^{\dagger}c_{\bm{k}'r\downarrow}
\!-\!c_{\bm{k}\ell\downarrow}^{\dagger}c_{\bm{k}'\ell\downarrow},&\quad
s_{+\downarrow}\!=\!s_{-\uparrow}^{\dagger}\!=\!c_{\bm{k}\ell\downarrow}c_{\bm{k}'r\downarrow}^{\dagger}.
\end{split}
\end{equation}
Note that $\mathcal{H}_{\rm 2CK}$ must be $\mathbb{T}$ invariant.
This is guaranteed since $S_i=-\mathbb{T}S_i$ following Eq.~(\ref{TF}) 
and $s_{i\bar\sigma}=-\mathbb{T}s_{i\sigma}$ $(i=x,y,z)$.
Moreover, the exchange couplings are derived to be
$J_z$$=$$(t_\ell^2+t_r^2) /m$,
$J_\pm$$=$$2t_\ell t_r/m$, and
$J_\delta$$=$$(t_\ell^2-t_r^2)/m$,
with $m^{-1}$$=$$(U_{2N\!+\!2}$$-$$U_{2N\!+\!1})^{-1}$$+$$(U_{2N}$$-$$U_{2N\!+\!1})^{-1}$$>$$0$.

Evidently, Eq.~(\ref{H2CK}) describes a 2CK model~\cite{Nozieres,Zawadowski,Affleck},  
in which the Kramers doublet of the floating TSC acts as a spin-$\frac{1}{2}$ impurity $\bm S$,
whereas the two spin species provide two independent reservoirs 
with left-right pseudospin $\bm{s}_\sigma$.
The left and right leads can flip $n_\ell$ and $n_r$, respectively, and hence flip $\bm S$. 
The channel symmetry in $J_{z,\pm}$ is guaranteed by $\mathbb{T}$ symmetry.
Because of the left-right anisotropy, a new exchange coupling $J_\delta$$\sim$$t_\ell^2\!-\!t_r^2$ appears.

The 2CK physics of the Majorana transistor is further illustrated in Fig~\ref{fig2}, 
compared with the 1CK physics of a singly occupied QD~\cite{GR,NL,Kastner,KG}.
In the first (last) two panels of Fig.~\ref{fig2}(a), 
the spin-$\downarrow$ ($\uparrow$) state of the QD
can be flipped in a co-tunneling event.
In the first (last) two panels of Fig.~\ref{fig2}(b),  
as one-to-one counterparts to those of Fig~\ref{fig2}(a), 
the spin-$\downarrow$ ($\uparrow$) state of the floating TSC 
can be similarly flipped in a co-tunneling event via the virtual state $(0,0,N)$ or $(1,1,N)$. 
In contrast to the QD case, these events only involve the electrons in the spin-$\uparrow$ channel.
As one-to-one $\mathbb{T}$ counterparts to those in Fig.~\ref{fig2}(b),
Fig.~\ref{fig2}(c) depicts the co-tunneling events in the spin-$\downarrow$ channel,
given that $\mathbb{T}$ flips spin, switches $n_{\alpha}$, but preserves $Q$.
Notably, the condensate must be involved here reflecting the superconducting nature,
and $\mathbb{T}$ symmetry dictates the channel symmetry. 

\begin{figure}[!t]
\includegraphics[width=1\columnwidth]{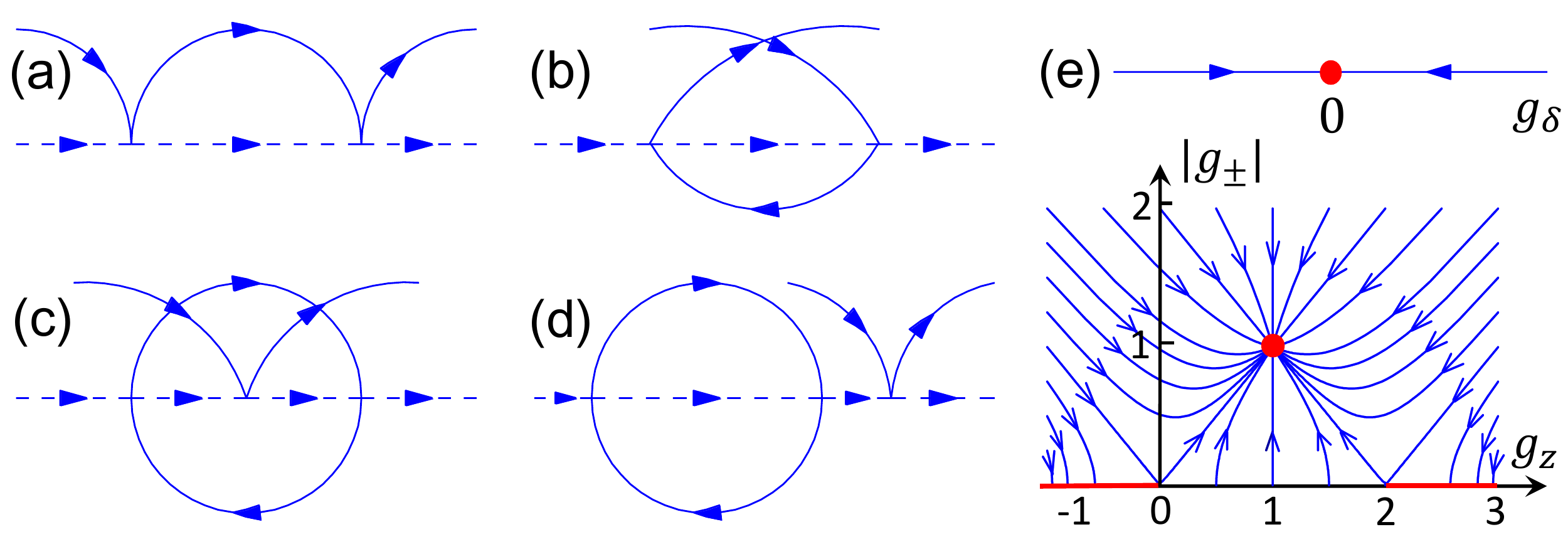}
\caption{(a)-(d) Sketches of the one- and two-loop diagrams for 2CK effects. 
(e) The RG flows of $g_\delta$, $g_{z}$, and $g_{\pm}$.}
\label{fig3}
\end{figure}

Now we use the perturbative renormalization group (RG) to analyze
the flows of exchange couplings under the influence of $J_{\delta}$. Computing the standard 
one- and two-loop Feynman diagrams~\cite{RG1,RG2,RG3,RG4,RG5,RG6} 
sketched in Figs.~\ref{fig3}(a)-(d), we derive the following flow equations:
\begin{eqnarray}
\begin{aligned}
\dv{g_{\delta}}{\ln\Lambda}=  g_{\pm}^{2}g_{\delta}\,,\;\;\;\;
\dv{g_{z}}{\ln\Lambda}= - g_{\pm}^{2} + g_{\pm}^{2} g_{z}\,,\\
\dv{g_{\pm}}{\ln\Lambda}= - g_{z}g_{\pm} + \frac{1}{2}\left(g_{\delta}^{2}+g_{z}^{2}+g_{\pm}^{2}\right)g_{\pm}\,,
\end{aligned}\label{PRG}
\end{eqnarray}
where $\Lambda$ is the momentum cutoff, $g_{\delta,z,\pm}\!=\!\nu J_{\delta,z,\pm}$ are dimensionless coupling constants,
and $\nu$ is the density-of-states of lead electrons.
When $g_{\delta}\!=\!0$, Eqs.~(\ref{PRG}) are exactly the same as those for the standard 1CK problem~\cite{RG2}
and can also be derived from the standard 2CK problem by enforcing a channel symmetry~\cite{Nozieres,Zawadowski,Affleck}. 
For the physically relevant regime $0\!<\!|g_{\pm,\delta}|\!<\!g_z\!\ll\!1$, as shown in Fig.~\ref{fig3}(e), 
$g_{\delta}$ flows to zero with vanishing left-right anisotropy,
whereas $(g_z, g_{\pm})$ flow to a stable intermediate-coupling fixed point $(1,1)$ with vanishing exchange anisotropy. 
Consequently, the ground state of the Majorana transistor
is a channel-symmetric, exchange-isotropic 2CK state. 
Integrating out Eq.~(\ref{PRG}) we further obtain $T_{\rm 2CK}\!\approx\!mge^{-1/g}$.
Consider an induced gap $\Delta\!\sim\! 1$~K ($10$~K) for an $s$-wave ($s_\pm$-wave) superconductor proximitized device
with $m\!\sim\! 0.3\Delta$, $T_{\rm 2CK}$ is estimated to be $30\!-\!150$~mK ($0.3\!-\!1.5$~K).

\indent\textcolor{blue}{\em Universal transport.}---The RG analysis becomes invalid upon approaching $T_{\rm 2CK}$; however, 
the conformal field theory~\cite{Affleck} has provided the scattering matrix $\mathcal{T}$,
which can be used to understand the universal transport mediated by the 2CK ground state.
The current in the $\alpha$ lead in Fig.~\ref{fig1}(b) is $I_{\alpha}\!=\!(ie/\hbar)[N_{\alpha}(t), \mathcal{H}]$,
which has equal contributions from the two channels because of $\mathbb{T}$ symmetry.
Following the standard derivation~\cite{Meir}, 
\begin{eqnarray}\label{Ialpha}
I_{\alpha}=\frac{2ie}{h}\int {\rm d}\omega\left[\Gamma_{\alpha}f_{\alpha}
(\mathcal{G}_{\alpha\alpha}^a-\mathcal{G}_{\alpha\alpha}^r)
-\Gamma_{\alpha}\mathcal{G}_{\alpha\alpha}^<\right]\,,
\end{eqnarray}
where $f_{\alpha}$ is the Fermi function, 
$\Gamma_{\alpha}\!=\!2\pi \nu t_{\alpha}^2$,
and $\mathcal{G}$ is the full Green's function.
Following the Dyson equation,
$\mathcal{G}_{\alpha\alpha}^a\!-\mathcal{G}_{\alpha\alpha}^r\!=
\mathcal{G}_{\alpha\alpha}^r(\Sigma_{\alpha\alpha}^a\!-\Sigma_{\alpha\alpha}^r)\mathcal{G}_{\alpha\alpha}^a$.
By the Keldysh equation, $\mathcal{G}_{\alpha\alpha}^<\!=\mathcal{G}_{\alpha\alpha}^r
\Sigma_{\alpha\alpha}^<\mathcal{G}_{\alpha\alpha}^a$.
In light of the fluctuation-dissipation theorem, 
$\Sigma_{\alpha\alpha}^<\!=\!f_{\alpha}(\Sigma_{\alpha\alpha}^a\!-\Sigma_{\alpha\alpha}^r)$.
Therefore, we conclude $I_{\alpha}\!=\!0$, no current across the TSC.

The vanishing of electron teleportation can be interpreted as follows.  
For each channel, electrons in the two leads have opposite left-right pseudospins 
that exchange couple to opposite spins of the floating TSC.
As the 2CK state and hence its $\mathcal{T}$ matrix~\cite{Affleck} are $SU(2)$ invariant,
any current in the leads would break the $SU(2)$ invariance and must be forbidden. 
The current also vanishes in the 
Coulomb blockade~\cite{KB} regime ($T_{\rm 2CK}\!<\!T\!\lesssim\!m$)
as the $\mathcal{T}$ matrix is diagonal in the pseudospin basis.

\begin{figure}[b!]
\includegraphics[width=0.95\columnwidth]{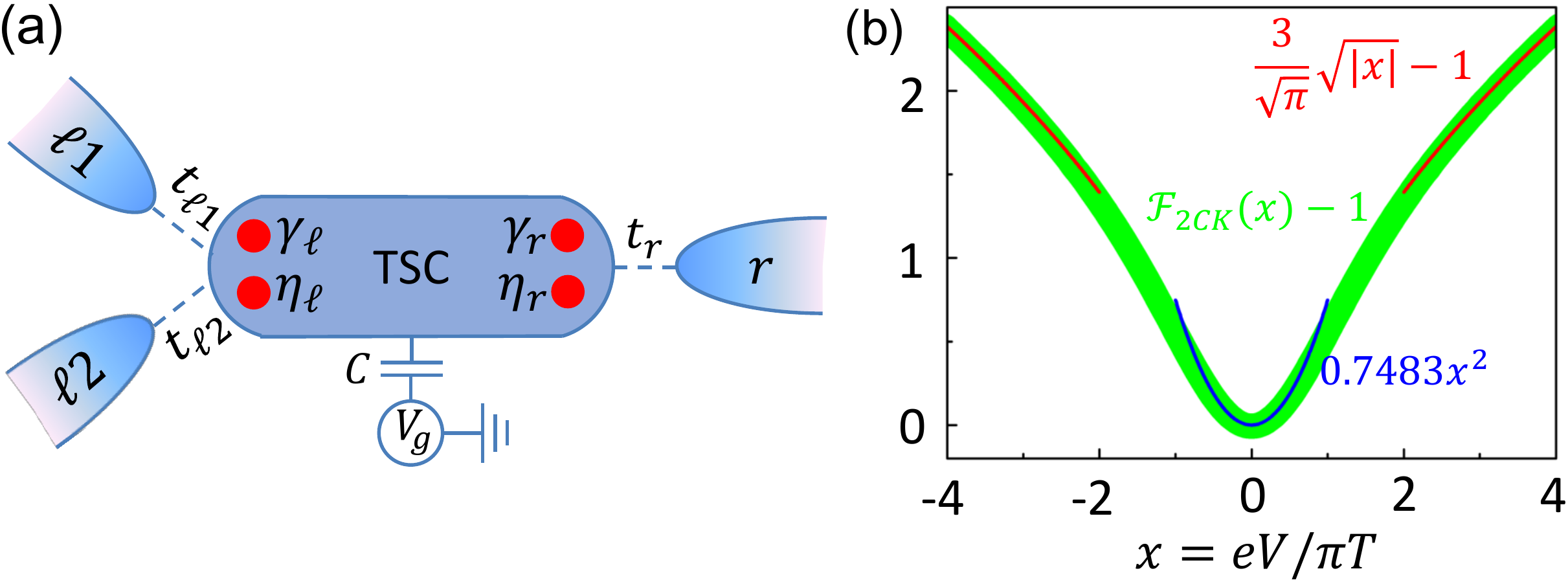}
\caption{(a) Schematics of a Majorana transistor similar to Fig.~\ref{fig1}(b) but with two split left leads. 
(b) The universal scaling curve $\mathcal{F}_{\rm 2CK}(eV/\pi T)-1$ 
and its asymptotes~\cite{ft1} for the differential conductance in a left lead of (a).}
\label{fig4}
\end{figure}

To reveal that the 2CK state can mediate nontrivial transport, 
we consider a three-terminal Majorana transistor sketched in Fig.~\ref{fig4}(a).
With a rotation in the space of the two left leads, 
$c_{\ell\sigma\pm}\!=\!(t_{\ell 1}c_{\ell 1\sigma}\pm t_{\ell 2}c_{\ell 2\sigma})/t_\ell$
with $t_\ell\!=\!(t_{\ell 1}^2+t_{\ell 2}^2)^{1/2}$,
Fig.~\ref{fig4}(a) can still be described by Eq.~(\ref{HT}) 
with $c_{\ell\sigma +}\!\to\! c_{\ell\sigma}$ and $c_{\ell\sigma -}$ decoupled.
We consider the current between the two left leads in the presence of the right lead.
By letting $\alpha\!=\!\ell 1,\ell 2$ in Eq.~(\ref{Ialpha}) 
and considering $I_{\ell}\!=\!I_{\ell 1}\!=\!-I_{\ell 2}$, we obtain the current~\cite{Meir}
\begin{eqnarray}\label{IL}
I_{\ell}=\frac{4e}{h}\int{\rm d}\omega\frac{\Gamma_{\ell 1}\Gamma_{\ell 2}}{\Gamma_{\ell 1}+\Gamma_{\ell 2}}
\left(f_{\ell 1}-f_{\ell 2}\right){\rm Im}\left[\mathcal{G}_{\ell\ell}^{r}\right]\,.
\end{eqnarray}
In the limit of $t_{\ell 2}\!\ll\!t_{\ell 1,r}$, the TSC will be in equilibrium with the $\ell 1$- and $r$-leads,
and the $\ell 2$-lead will only probe the TSC even at finite bias $V$. 
As $\mathcal{G}_{\ell\ell}^{r}\!=\!\mathcal{T}/t_{\ell}^2$ 
and $G_{\ell}\!=\!{\rm d}I_{\ell}/{\rm d}V$, the conductance (for $T,|eV| < T_{\rm 2CK}$) reads
\begin{eqnarray}\label{GV}
G_{\ell}(V,T)=G_{0}\left[1-\sqrt{\frac{\pi T}{T_{\rm 2CK}}}\mathcal{F}_{\rm 2CK}\left(\frac{eV}{\pi T}\right)\right]\,,
\end{eqnarray}
where $\mathcal{F}_{\rm 2CK}(0)\!=\!1$ and $G_{0}\!=\!4t_{\ell 1}^2t_{\ell 2}^2/t_{\ell}^4\cdot e^2/h$. 
Strikingly,
$[G_{\ell}(0,T)-G(V,T)] G_0^{-1} \sqrt{T_{\rm 2CK}/\pi T}$ is a universal scaling function:
$\mathcal{F}_{\rm 2CK}(eV/\pi T)-1$~\cite{ft1}, as plotted in Fig.~\ref{fig4}.

To shed more light on how the right lead telecontrols the transport through the left end, 
we contrast Fig.~\ref{fig4}(a) with a double-QD device~\cite{OGG1,OGG2,OGG3},
where Oreg and Goldhaber-Gordon have ingeniously realized a 2CK effect.
In their case, the 2CK transport occurs only in one spin-degenerate channel, 
under a nonlocal influence of the other channel (i.e. the larger QD);
when the larger QD is decoupled, a 1CK effect returns.
In our case, the 2CK transport occurs in both channels but only in the left leads (i.e. pseudospin polarized),
under the nonlocal influence of the right lead at $T\!<\!T_{\rm 2CK}$;
when the right lead is decoupled, there are no Kondo effects:
only co-tunnelings on top of the Coulomb blockade~\cite{Marcus,KB} at $T\!\lesssim\!m$.
In both cases, there is one decoupled electron channel,
and this highlights that Fig.~\ref{fig1}(b) utilizes 
the minimum number of leads to realize the 2CK effect.

\indent{\color{blue}\em Discussions.}---It is instructive to consider the consequences of symmetry breaking in our Majorana transistor.
(i) If $\mathbb{T}$ symmetry is only broken in a ferromagnetic lead,  
the two channels become asymmetric, 
and the system flows to the 1CK fixed point in the more strongly coupled channel.
(ii) If spin-orbit couplings are substantial in a lead even if $\mathbb{T}$ symmetry is unbroken, 
the two channels are not independent, and the system also flows to the 1CK fixed point.
(iii) If $\mathbb{T}$ symmetry is broken in the TSC, there are two scenarios. 
The SC may still be topological with an unpaired Majorana left at each end~\cite{FZ1}. 
Whereas a 2CK effect is impossible, a charge 1CK effect could occur 
when the charge degeneracy point (i.e. $Q_0/e$ is a half integer) 
is fine-tuned~\cite{Charging,BB}, as considered by Fu.
If the SC is trivial, each Majorana Kramers pair splits into a pair of Andreev states 
with opposite energies. With left-right anisotropy,
this produces a Zeeman splitting between the $(0,1,N)$ and $(1,0,N)$ states, 
driving a Fermi-liquid crossover~\cite{Affleck,Pustilnik}.

There exist two seminal 2CK models with broken $\mathbb{T}$ symmetry. 
Fiete {\it et al.} proposed a model by  
coupling a level-$k$ Read-Rezayi fractional quantum Hall state at $\nu\!=\!2\!+\!k/(k\!+\!2)$ 
to a QD at a fine-tuned charge degeneracy~\cite{Fiete}. 
Here a Majorana is central to the $\nu\!=\!5/2$ case. 
B\'eri-Cooper and Altland-Egger constructed a model by coupling $M\!\geq\!3$ Majoranas 
of multiple class-D TSCs to $M$ leads~\cite{Beri1,Egger,Beri2,Mich,Mora}.
The $M\!=\!4$ case is a 2CK model~\cite{Egger,Beri2},
and the Majorana hybridization is a relevant perturbation destabilizing the 2CK fixed point~\cite{Beri3}.

By contrast, the setup of our $\mathbb{T}$-invariant model is extremely neat;
neither the channel symmetry nor the charging energy requires fine-tuning.
Strikingly, our 2CK effect is robust against the Majorana hybridization 
as long as the splitting is smaller than $e^2/C$,
because the topological degeneracy between the $(0,1,N)$ and $(1,0,N)$ states 
is also a Kramers degeneracy of a many-body state with an odd number of fermions. 
Hence, the $\mathbb{T}$ symmetry is indeed superior, 
given that the channel symmetry is notoriously difficult to realize 
and the Majorana hybridization is inevitable in practice.
Whereas we have chosen the odd-parity degeneracy above, 
there is a dual 2CK effect for the even-parity degenerate states $(0,0,N\!+\!1)$ and $(1,1,N)$
when $Q_0/e$ is rounded to $2N\!+\!2$. 
The even-parity problem can be mapped to the odd-parity one 
by redefining $d_{\alpha}$$=$$(\gamma_{\alpha}$$+$$i\alpha\eta_{\alpha})/2$ with $\alpha\!=\!\ell/r\!=\!\pm$,
yet the even-parity degeneracy is not a Kramers degeneracy and not robust to the Majorana hybridization.

It is intriguing to consider the case in which the TSC is grounded ($C\!=\!\infty$). 
Without the charging energy, the two tunnelings in a co-tunneling event is no longer coherent, 
and Andreev reflections (AR) violating charge conservation is energetically favored.
Consider the Landauer formula for SCs~\cite{Datta} with the $\mathcal{S}$ matrix
\begin{eqnarray}
S(E)=1-2\pi i\mathcal{W}^{\dag}(E+i\pi \mathcal{W}\mathcal{W}^{\dagger}-\mathcal{H}_M)^{-1}\mathcal{W}\,,
\end{eqnarray} 
where $\mathcal{W}$ is a matrix describing the couplings of the Majoranas to the leads,  
as implied by Eq.~(\ref{HT}), and $\mathcal{H}_M\!=\!0$ for unhybridized Majoranas. 
At zero bias ($E\!\to\! 0$), this yields $G_\alpha\!=\!4e^2/h$ 
reflecting resonant local AR and vanishing cross AR at each channel~\cite{LAR},
consistent with a previous prediction~\cite{FZ1}.
If the left and right Majoranas were hybridized due to a finite size effect, 
the crossed (local) AR would be enhanced (suppressed)~\cite{CAR}. 

Finally, the Majorana transistor offers 
a unique platform for exploring strongly correlated physics,
spotlighting the Majorana-Majorana interactions.
The striking 2CK effect and their reductions in various limits can also provide 
fingerprints of Majorana Kramers pairs in their experimental detections. 
In the experiment coupling the Josephson junction and the 2D TI~\cite{Knez,Hart,Shi,Pri,Boc},
the phase difference $\delta\phi$ can be controlled by a magnetic flux. 
If the junction is weakly probed at the two edges by the source and drain,
the studied 2CK effect should be realized at $\delta\phi\!=\!\pi$,
and the discussed Zeeman-splitting scenario should apply to $\delta\phi\!\neq\!\pi$.
In the experiment coupling the Rashba nanowire and the $s_\pm$-wave SC~\cite{FZ1},
the topological phase transition can be tuned by a gate voltage.
For a TSC, the Majorana Kramers pairs should mediate the studied 2CK effect.
For a trivial SC, the Coulomb blockade with an even-odd effect should occur 
due to the charging energy and the quasiparticle gap~\cite{Marcus,final}. 

\indent{\color{blue}\em Acknowledgments.}---
F.Z. thanks Charles Kane, Liang Fu, Ion Garate, and Yichen Hu for insightful discussions.


\begin{thebibliography}{100}
\bibitem{ft1} $\mathcal{F}_{\rm 2CK}(x)$$=$$\int_{0}^{1}{\rm d}u\left[\frac{3}{\pi\sqrt{u(1-u)^3}}-
\frac{3F(u)}{4}\cos\left(\frac{x}{2}\log u\right)\frac{|\log u|}{\sqrt{1-u}}\right]$,
where $F(u)\!=\!F(\frac{3}{2},\frac{3}{2};1;u)$ is hypergeometric function.
$\mathcal{F}_{\rm 2CK}(x)$~is~$0.7483x^{2}\!+\!1$ for $|x|\!\ll\!1$ and $3\sqrt{\frac{|x|}{\pi}}$ for $|x|\!\gg\!1$.
\end{thebibliography}
\end{document}